\documentclass[aps,prb,twocolumn,floatfix,floats]{revtex4}
\usepackage{graphicx,latexsym}

\begin{document}

\title{Impurity and spin effects on the magneto-spectroscopy \\ of a
       THz-modulated nanostructure}

\author{Vidar Gudmundsson}
\altaffiliation[Permanent address: ]{Science Institute, University of Iceland, 
        Dunhaga 3, IS-107 Reykjavik, Iceland}
\affiliation{Physics Division, National Center for Theoretical Sciences,
        P.O.\ Box 2-131, Hsinchu 30013, Taiwan}
\author{Chi-Shung Tang}
\affiliation{Physics Division, National Center for Theoretical Sciences,
        P.O.\ Box 2-131, Hsinchu 30013, Taiwan}
\author{Andrei Manolescu}
\affiliation{Science Institute, University of Iceland, 
             Dunhaga 3, IS-107 Reykjavik, Iceland}

%
%

\begin{abstract} 
We present a grid-free DFT model appropriate to
explore the time evolution of electronic states 
in a semiconductor nanostructure. The model can be used to 
investigate both the linear and the nonlinear response of the system
to an external short-time perturbation in the THz regime.
We use the model to study the effects of impurities on 
the magneto-spectroscopy of a two-dimensional electron gas in a 
nanostructure excited by an intense THz radiation. 
We do observe a reduction in the binding energy of the impurity
with increasing excitation strength, and at a finite magnetic
field we find a slow onset of collective spin-oscillations
that can be made to vanish with a stronger excitation.
\end{abstract}


\maketitle
%
\section{Introduction}
The interest in the properties of radiation induced excited states in
bulk semiconductors,\cite{Plancken95:9643,Nie03:1}  
or a two-dimensional electron gas (2DEG),\cite{Mani02:646}
and semiconductor nanostructures\cite{Lorke00:2223} 
has grown in the last years parallel to 
advances in high precision experimental techniques and sample making skills.
Increased brilliance of coherent far-infrared (THz) synchrotron 
radiation\cite{Abo-Bakr03:094801} and sensitive photoconductivity 
measurements techniques\cite{Plancken95:9643,Nie03:1} will certainly intensify 
research activities on electronic systems driven far from
equilibrium and time-dependent phenomena in nanostructures.

We model and investigate here the time evolution of a short nanoscale quantum 
wire or an elongated quantum dot with one embedded impurity after the
system with several electrons has been irradiated by a short intense  
THz pulse. Corresponding atomic systems have been 
investigated,\cite{Chu01:023411} but the present model of an artificial
two-dimensional (2D) molecule in a GaAs structure allows us to consider 
the effects of an external magnetic field $B<2$ T, that is commonly
available in the labs and has large effects on the 2DEG in a AlGaAs/GaAs
heterostructure, but very small effects on usual atomic systems.

Having seen that an intensive THz pulse can cause nonlinear effects
in quantum dots\cite{Puente01:235324} and rings\cite{Gudmundsson03:161301} 
and possibly change the binding
energy of impurities in bulk material\cite{Nie03:1} we study its effects
in the present system on the binding energy of the impurity in presence
of other interacting electrons, and ask ourselves if it is possible
to excite collective spin modes in the system in an external
magnetic field. THz radiation with the equivalent energy of few meV is simply on an 
appropriate scale with the energy spectrum of our system determined
by its geometry (finite size) and the host material of GaAs. 

Modeling time-dependent phenomena in nanosystems in a magnetic field 
driven out of equilibrium does not leave many options to a researcher.
The Coulomb interaction between the electrons is usually of paramount
interest together with the spin structure of the system. This can be
emphasized by its finite size even though the effective g-factor may be 
small as is the case in GaAs. The well established time-dependent Hartree-Fock 
approximation can be adequate in many cases,\cite{Puente01:235324} but
one would hope for more accurate spin structure delivered by a 
time-dependent local spin-density approximation (LSDA).\cite{puente99:3266}
It has to be kept in mind that both the inclusion of the magnetic field
in a LSDA\cite{Qian03:066402} and its extensions to dynamic 
systems\cite{Aryasetiawan02:165119} are nontrivial tasks and are not fully 
completed to date. In our approach we shall use the ground state 
parameterization of the exchange and correlation functionals and ignore
the small corrections due to current terms in the functionals in a
magnetic field.\cite{Vignale87:2360} 

Commonly, both in calculations for excited 
atoms and molecules\cite{Bandrauk00:053406,Chu01:023411,Chu01:063404} 
or nanostructures,\cite{puente99:3266,Puente01:235324}
a spatial grid has been employed. In this work we resort to
recently developed grid-free implementation of 
LSDA\cite{Zheng93:397,Glaesemann98:9959,Glaesemann99:6580,Berghold98:344} in the
hope to be able to describe complex dynamical systems for a longer period of
time. The grid-free methods, as will become clear below, result in the 
transformation of the relevant equations to compact matrix equations
of not too large size. On the other hand the matrix elements are usually
much more complicated than in grid methods, even though most of them
are known analytically. This compact matrix structure of the problem
lends itself very well to the matrix and vector handling of modern
computer languages and facilitates parallelization that does neither require
lare bandwidth nor high communication speed between the nodes of a CPU cluster.    

%

\section{Model}\subsection{The ground state}
To calculate the time evolution of the system requires the knowledge
of the initial state, its ground state, at a particular time $t=t_0$,
which within the LSDA is described by the Kohn-Sham equations
\begin{equation}
      H|\alpha ) = \left(  H_0 + H_\sigma + V_{\phi} + H_{\mathrm{int}} 
      \right) |\alpha ) = \varepsilon_\alpha |\alpha ) ,
      \label{Kohn_Sham} 
\end{equation}
where $H_0$ is the Hamiltonian for the 2DEG in the perpendicular 
magnetic field ${\bf B}=B{\bf\hat z}$, that is confined by the
parabolic potential $V_{\mathrm{conf}}(r)=m^*\omega_0^2r^2/2$.
A more general shape of the system is attained through the 
additional confinement potential
$V_{\phi}$ breaking the rotational symmetry of $H_0$ and modifying 
the radial confinement
\begin{equation}
      V_{\phi}({\bf r}) = \frac{1}{2}m^*\omega_0r^2
      \sum_{p=1}^{p_{\mathrm{max}}}\alpha_p\cos{(p\varphi )}
      + V_0\exp{(-\gamma r^2)},
\label{V_c_i}
\end{equation}
where $p_{\mathrm{max}}$ is an even integer to ensure confinement.
The eigenstates $|\alpha\rangle$ of the Fock-Darwin\cite{Fock28:446}
Hamiltonian $H_0$ are used as a mathematical basis in which the eigenstates
$|\alpha )$ of $H$ are expanded, as a result Eq.\ (\ref{Kohn_Sham}) can be transformed
into a matrix eigenvalue equation. A suitable truncation of the
basis $\{|\alpha\rangle\}$, that can be varied to improve the accuracy,
then leads to a numerically tractable eigenvalue problem.

In this LSDA mean-field type of an approximation the effective interaction
$H_{\mathrm{int}}$ is composed of the direct Coulomb repulsion potential,
or the Hartree term
\begin{equation} 
      V_{\mathrm{H}}({\bf r})=\frac{e^2}{\kappa}
      \int \frac{n({\bf r}')}{|{\bf r}-{\bf r}'|}d{\bf r}',
\end{equation}
and the potential describing the effects of the exchange and 
correlation
\begin{equation}
       V_{xc,\sigma}(r,B)=\frac {\partial}{\partial n_{\sigma}}
      (n\epsilon_{xc}[n_{\uparrow},n_{\downarrow}, B])
      |_{n_{\sigma}=n_{\sigma}(r)},
      \label{V_xc}
\end{equation}
where $n_{\sigma}(r)$ is one of the two spin ($\sigma =\uparrow ,\downarrow$) 
components of the total electron density 
$n = n_{\uparrow} + n_{\downarrow}$. The Zeeman energy of the electrons
is included in $H_\sigma$ with the appropriate effective g-factor $g^*$
and the Bohr magneton, $\mu_B$, $H_\sigma =\pm(1/2)g^*\mu_B B$.

In a large 2DEG the natural length and energy scales are the magnetic
length $l=\sqrt{\hbar c/(eB)}$ and the cyclotron energy 
$\hbar\omega_c=\hbar eB/(m^*c)$, and instead of the electron density it is 
convenient to define the local filling factor  
$\nu ({\bf r}) = 2\pi l^2n({\bf r})$. In the confined system described
by $H_0$ the appropriate length scale is $a = l\sqrt{\omega_c/\Omega}$,
an ``effective magnetic length'', where $\Omega = \sqrt{\omega_c^2 + 4\omega_0^2}$ 
is the new characteristic frequency of the system. We shall thus use a 
modified filling factor appropriate for a finite system
\begin{equation}
      \tilde\nu ({\bf r}) = 2\pi a^2n({\bf r}), 
\end{equation}
and the polarization $\zeta({\bf r}) = 
[n_{\uparrow}({\bf r}) - n_{\downarrow}({\bf r})]/n({\bf r})$
as new variables to express the exchange and correlation potentials
(\ref{V_xc}) in. This will allow us to have a unified description
of the system in a finite or vanishing magnetic field.

In these new variables the exchange and correlation potentials 
are\cite{Lubin97:10373}
\begin{eqnarray}
       V_{xc,\uparrow}=\frac {\partial}{\partial \tilde\nu}
      (\tilde\nu \epsilon_{xc})+(1-\zeta)\frac{\partial}
      {\partial \zeta}\epsilon_{xc}\nonumber\\
      V_{xc,\downarrow}=\frac {\partial}{\partial \tilde\nu}
      (\tilde\nu \epsilon_{xc})-(1+\zeta)\frac {\partial}
      {\partial \zeta}\epsilon_{xc} .
\end{eqnarray}
In the magnetic field the exchange and correlation energy density is
interpolated between the infinite and the no-field limits 
as\cite{Koskinen97:1389}
\begin{eqnarray}
      \epsilon_{xc}^B(\tilde\nu , \zeta)=\epsilon_{xc}^{\infty}
      (\tilde\nu)e^{-f(\tilde\nu)}+\epsilon_{xc}^0 
      (\tilde\nu, \zeta)(1-e^{-f(\tilde\nu)}),
\end{eqnarray}
with $f(\tilde\nu )= (3\tilde\nu /2) + 7\tilde\nu^4$ and
$\epsilon_{xc}^{\infty}(\tilde\nu)=-0.782\sqrt{\tilde\nu}e^2/(kl)$.
The low-field energy functional is interpolated between 
the spin-polarized and the unpolarized limits\cite{Barth96:8411}
\begin{eqnarray}
      \epsilon_{xc}^0(\tilde\nu,\zeta)=\epsilon_{xc}(\tilde\nu,0)+f^i(\zeta) 
      \left[ \epsilon_{xc}(\tilde\nu,1)-\epsilon_{xc}(\tilde\nu,0) \right]\nonumber\\
      \mbox{with}\quad
      f^{i}(\zeta)=\frac {(1+\zeta)^{3/2}+(1-\zeta)^{3/2}-2}{2^{3/2}-2}.
\end{eqnarray}
The exchange and correlation contributions are then separated 
$\epsilon_{xc}(\tilde\nu,\zeta)=\epsilon_x(\tilde\nu, \zeta)
+ \epsilon_c(\tilde\nu,\zeta)$, 
and the exchange contribution for the unpolarized system is expressed as 
$\epsilon_x(\tilde\nu,0)=-[{4}/(3\pi)] 
\sqrt{\tilde\nu}{e^2}/(kl) $, while for the polarized one it is 
$\epsilon_x(\tilde\nu,1)=-[{4}/(3\pi)] \sqrt{2\tilde\nu} e^2/(kl)$. 
We use the parameterization of Ceperley and Tanatar for the 
correlation contribution\cite{Tanatar89:5005}
\begin{eqnarray}
       \epsilon_c(\tilde\nu,\zeta)=a_0 \frac {1+a_1 x}{1+a_1 x+a_2 x^2+a_3 x^3}
      E_{Ryd}^{\ast},
\end{eqnarray}
where $x=\sqrt{r_s}=( 2 /\tilde\nu )^{1/4}( a / a_B^{\ast})^{1/2}$,
and $a_B^{\ast}$ is the effective Bohr radius. The optimized values
of the correlation parameters $a_i$ have been found by Ceperley and Tanatar
using a Monte Carlo model of a 2DEG.\cite{Tanatar89:5005}

So stated, the $V_{xc,\sigma}$'s are complicated functionals of the
the spin densities, or $\tilde\nu$ and $\zeta$, and thus a common method
is to solve the resulting Kohn-Sham equations on a spatial grid.
We instead follow the alternative example of a number of researchers to use the
mathematical basis $\{ |\alpha\rangle\}$ to cast the LSDA functionals
into matrix expressions in order to implement a grid-free 
LSDA.\cite{Zheng93:397,Glaesemann98:9959,Glaesemann99:6580,Berghold98:344}
Briefly, the method relies on the following steps: In the end we need
the matrix elements of the exchange and correlation potentials,
$\langle\alpha |V_{\mathrm{xc,\sigma}}|\beta\rangle$, so we start by 
constructing $\langle\alpha |\tilde\nu |\beta\rangle$. The polarization
$\zeta$ has to be treated as a combination of two functions,
$f = \tilde\nu_\uparrow - \tilde\nu_\downarrow$ and 
$g = \tilde\nu^{-1}$. Due to the completeness of the basis we then have
$\langle\alpha |\zeta|\beta\rangle =
\sum_\gamma\langle\alpha |f |\gamma\rangle\langle\gamma |g|\beta\rangle$,
and to evaluate the latter matrix elements we need to find a unitary
matrix ${\bf U}$ that diagonalizes the matrix $\mathbf{\tilde\nu}$,
i.e.\, $\mathbf{\tilde\nu} = 
{\bf U}\;\mathrm{\mathbf{diag}}(\lambda_1,\cdots,\lambda_n)
{\bf U}^{+}$, where $\lambda_i$ are the eigenvalues of $\mathbf{\tilde\nu}$.
All functions of the matrix $\mathbf{\tilde\nu}$ are now calculated
according to
\begin{equation}
      f[\mathbf{\tilde\nu}] = {\bf U}\;\mathrm{\mathbf{diag}}
      \left(f(\lambda_1),\cdots,f(\lambda_n)\right){\bf U}^{+}. 
\end{equation}
Corresponding methods are used for more complex functionals of 
$\tilde\nu$ and $\zeta$. 

The matrix elements of the electron density $n$ or the effective
filling factor $\tilde\nu$ can be written in terms of a 
generalized one-electron overlap integral and the matrix elements
of the density operator $\mathbf{\rho}$
\begin{equation}   
      \langle\alpha |\tilde\nu |\beta\rangle = \sum_{p,q}
      \rho_{qp}\int d\mathbf{r}\;\phi_\alpha^*(\mathbf{r})
      \phi_p^*(\mathbf{r})\phi_q(\mathbf{r})\phi_\beta (\mathbf{r}).
\end{equation}
In the ground state the density matrix is constructed from the
expansion coefficients of electronic states in terms of the basis
states, the occupation of the states expressed by the
Fermi distribution, and the chemical potential $\mu$ ensuring the
conservation of the number of electrons, 
$\rho_{qp} = \sum_\gamma f(\varepsilon_\gamma -\mu )
c^*_{\gamma p}c_{\gamma q}$.
We use here the density matrix $\rho_{qp}$ in anticipation of the situation
when the system will be subjected to a strong external time-dependent 
perturbation and quantities like the single-electron energy spectrum,
and the occupation described by the Fermi distribution will have no
meaning. The analytic expression for the generalized one-electron 
overlap integral can be found in Appendix. The matrix elements of the
direct Coulomb interaction and the confinement potential have been
published elsewhere.\cite{Ingibjorg99,Vasile02:hi}

\subsection{Time-dependent excitation}
At $t=t_0$ the Hamiltonian of the system acquires time dependence,
$H(t)=H+W(t)$, caused by an external perturbation of a finite duration
\begin{eqnarray} 
      W(t) &=& V_t r^{|N_p|}\cos{(N_p\phi)}\exp{(-sr^2-\Gamma t)}\nonumber\\ 
           &{\ }&\sin{(\omega_1t)}\sin{(\omega t)}\theta (\pi -\omega_1t) .
\label{Wt}
\end{eqnarray}
Since the perturbation is of arbitrary strength we have to assume the
task of solving the equation of motion for the density operator
\begin{equation}
      i\hbar d_t{\rho}(t) = [H + W(t),\rho (t)] .
\end{equation}
The structure of this equation is inconvenient for numerical evaluation
so we resort instead to the time-evolution operator $T$, defined by
$\rho (t) = T(t)\rho_0T^+(t)$, which has simpler equation of motion
\begin{eqnarray}
      i\hbar\dot T(t)   &=& H(t)T(t)\\  \nonumber    
     -i\hbar\dot T^+(t) &=& T^+(t)H(t)  .
\label{Teq}     
\end{eqnarray}
In each time step the density operator $\rho$ changes and thus also
the DFT Hamiltonian which is a functional of $\rho$. Similar nonlinear
problem is encountered when the time-dependent Kohn-Sham equations 
are solved directly on a spatial grid.\cite{puente99:3266}
We apply a similar procedure here discretizing time and using the
Crank-Nicholson algorithm for the time integration resulting in the
coupled equations
\begin{eqnarray}
\label{CN-t}  
      &{\ }&\rho(t+\Delta t) = T(\Delta t)\rho(t)T^+(\Delta t)\\ \nonumber
       &{\ }&\left\{1+\frac{i\Delta t}{2\hbar}H[\rho ;t+\Delta t]
      \right\}T(\Delta t) \approx
      \left\{1-\frac{i\Delta t}{2\hbar}H[\rho ;t]\right\} ,  
\end{eqnarray}
with the initial condition, $T(0)=1$, already used on the right hand
side of (\ref{CN-t}). In each time-step the equations have to be iterated
until self-consistency is attained. This is performed in the truncated
Fock-Darwin basis $\{|\alpha\rangle\}$ for the corresponding
matrix version of the equations (\ref{CN-t}). 

The accuracy and the fine-tuning of the model has been tested for the
ground state with comparison to the exact results calculated for a 
circular parabolically confined quantum dot,\cite{Pfannkuche93:2244} 
and in the time-evolution by convincing us that the model can reproduce
the Kohn results for the absorption of a Far-infrared 
radiation.\cite{Maksym90:108,Gudmundsson91:12098} In addition, the
normalization condition for the density matrix is tested regularly
during the integration of the time-dependent problem.

\subsection{Specification of model parameters}
\begin{figure}[htbp!] 
\begin{center}
\includegraphics[width=6.0cm]{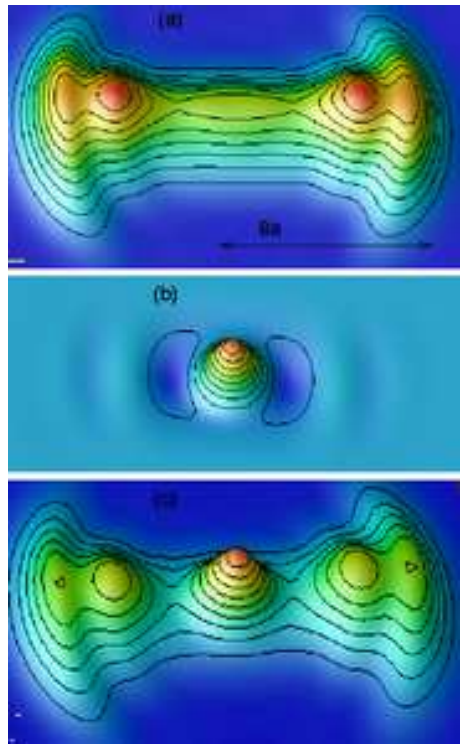}
\end{center}
\caption{Color online. 
         The electron density of the interacting system without an impurity
         (a), the difference in the electron density of system
         with and without an impurity (b), and the density for a system
         with an impurity (c). $B=0$ T, $V_0=-10$ meV, $T=1$ K, $N=10$,
         and $a\approx 13$ nm.}
\label{Density}
\end{figure}

\begin{figure}[htbp!] 
\begin{center}
\includegraphics[bb=170 270 420 560,clip,width=8.0cm]{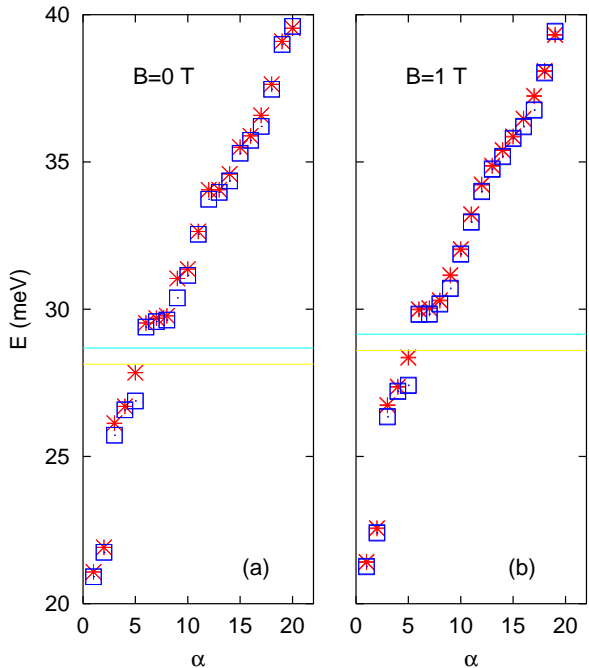}
\end{center}
\caption{Color online.
         The Kohn-Sham single-electron energy spectra for the 
         interacting electron system with ($\Box$) 
         and without ($*$) an impurity for $B=0$ T (a), and
         $B=1$ T (b). The integer $\alpha$ is a quantum number 
         labeling the states. The lower horizontal
         line (yellow) represents the chemical potential $\mu$ for
         system with an impurity, and the higher one (blue) indicates
         $\mu$ for a system without an impurity. 
         $T=1$ K and $N=10$.}
\label{Espectrum}
\end{figure}

In order to model a short GaAs quantum wire with or without an embedded
impurity and several electrons we select the parameters, 
$\hbar\omega_0 = 3.37$ meV, and assume $g^* = 0.44$, and $m^* = 0.067m$. 
The impurity potential is selected to be Gaussian (\ref{V_c_i})
with $V_0 = -10$ meV, and $a^2\gamma =1.0$. The Gaussian shape is 
used here in order to limit the size of the basis during the 
CPU-intensive time integration of the problem.
The wire shape of the system is attained by setting
$\alpha_2=-0.7$ and  $\alpha_i=0$ for all $i\neq 2$.
The charge density for 10 electrons in the system is
shown in Fig.\ \ref{Density} with and without an embedded
impurity. At $B=0$ T the length of the system is approximately
200nm, and the difference between the charge densities with and 
without an impurity in the system turns up in Fig.\ \ref{Density}b
as a fairly well localized impurity state. 
%
%
%
%
This can be confirmed by the Kohn-Sham energy spectra displayed in
Fig.\ \ref{Espectrum}, where the main change due to the addition of
the impurity potential into the system is the lowering of the
energy of the highest almost degenerate doubly occupied state by approximately
1 meV.  
%
%
%
%
The calculations reported here are for $B=0$ T, and
$B=1$ T. In order to avoid numerical problems with the 
determination of the eigenvectors for exactly degenerate
eigenvalues in the diagonalization of the Hamiltonian we
actually use $B=10^{-8}$ T instead of $B=0$ T.

For the radiation pulse (\ref{Wt}) of duration just over 3 ps,\cite{Plancken95:9643} 
we select $\Gamma = 2$ ps$^{-1}$.
The envelope frequency corresponds to
$\hbar\omega_1 = 0.658$ meV, and the base frequency is equivalent to
$\hbar\omega = 2.63$ meV.
Meanwhile, we select a $|N_p|=1$ and $s=0$ in (\ref{Wt})
to represent a dipole radiation.
But, due to the finite cut-off in
the basis $|\alpha\rangle$ the matrix elements of the 
excitation can never be purely of the dipole type. 
We choose the orientation of the initial radiation pulse
(\ref{Wt}) to be either parallel, $\parallel$, to the long axis of the
system or perpendicular to it, $\perp$.  
The finite asymmetric cut-off of the basis can introduce a tiny 
mixing of the perpendicular and the parallel excitation at 
vanishing magnetic field. In fact, the basis cut-off is
performed by specifying the highest radial quantum number
$n_{\mbox{max}}$ used in the calculation, by doing so
an asymmetry is introduced between the number of negative and
positive angular quantum numbers $M$. This asymmetry is
made unimportant with respect to the energy spectrum and
the wave functions of the system by selecting a high enough
$n_{\mbox{max}}$, but the representation of a dipole potential
is not quite so clean. A simple alternative method is not
easy since the effective asymmetry does depend on the magnetic
field.   
  
We integrate the equations for the time-evolution operator
(\ref{Teq}) over a time interval of 50 ps using an increment
$\Delta t = 0.0025$ ps and evaluate the expectation values 
for the coordinates $x$ and $y$ for each spin direction, 
$\langle x\rangle_\uparrow$, etc. We can thus identify
oscillations perpendicular or parallel to the long axis
of the system, talking about perpendicular or parallel
detection. Resonances in the collective
oscillations of the electron spin densities are sought by 
Fourier transformation of the time series in the 
interval 5 $-$ 50 ps, when the external excitation has been 
turned off and switch-on effects have vanished and the system
has reached a ``steady state''.

\section{Results}     
\subsection{The $B=0$ T case}
\begin{figure}[htbp!]
\begin{center}
\includegraphics[bb=152 215 430 615,clip,width=8.0cm]{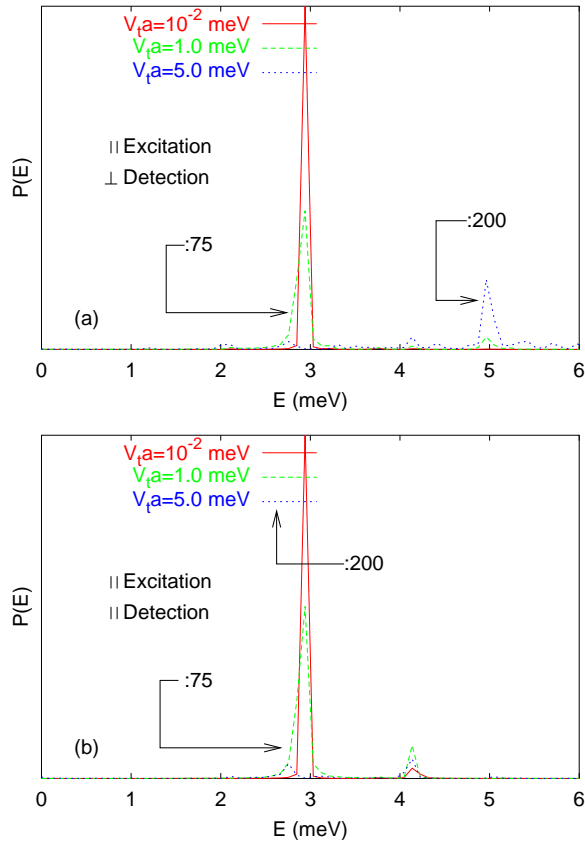}
\end{center}
\caption{Color online.
         The Fourier power spectrum of the time series for the
         expectation values of the coordinates 
         perpendicular ($\perp$) (a),
         and parallel ($\parallel$) (b) to the long axis 
         of the system. The excitation is {\bf\em parallel} to the same
         axis. Division factors for the stronger excitations
         are indicated in the figure. {\bf\em No impurity} is in the
         system. $B=0$ T, $T=1$ K, and $N=10$. }
\label{P0y}
\end{figure}

\begin{figure}[htbp!] 
\begin{center}
\includegraphics[bb=152 215 430 615,clip,width=8.0cm]{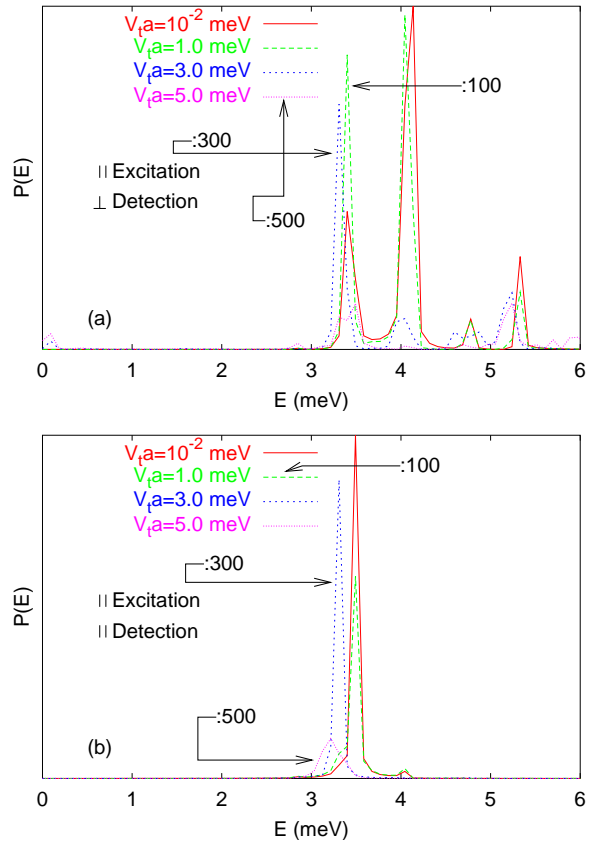}
\end{center}
\caption{Color online.
         The Fourier power spectrum of the time series for the
         expectation values of the coordinates 
         perpendicular ($\perp$) (a),
         and parallel ($\parallel$) (b) to the long axis 
         of the system. The excitation is {\bf\em parallel} to the same
         axis. Division factors for the stronger excitations
         are indicated in the figure. {\bf\em Impurity is present}
         in the system. $B=0$ T, $T=1$ K, and $N=10$. }
\label{Piy}
\end{figure}

In Fig.\ \ref{P0y} the Fourier power spectra of the time series for
the coordinates $x$ and $y$ are used to represent the excitation 
spectra of the system without an impurity. The $\parallel$ polarized
excitation is seen to introduce a tiny $\perp$ oscillation 
($1/10$ the amplitude) as has been explained earlier, 
but at the same frequency. This simple excitation
spectrum does not change much with increasing excitation 
strength $V_t$. A second harmonic shows up in the parallel 
oscillation, but more interestingly the 
height of the main peak does not grow linearly with the excitation 
strength when $V_ta>1$ meV. The excitation then becomes nonlinear and
as $V_ta=5$ meV we see that it shifts slightly to lower energy.      
%
%
%
%
Before commenting on the red shift of the peak with increasing $V_t$
we show what happens in the system with an embedded impurity in 
Fig.\ \ref{Piy}. The main peak in the longitudinal oscillation 
(Fig.\ \ref{Piy}b) is slightly blue shifted compared to the system with 
no impurity. The reason is readily explained by a glance at the 
Kohn-Sham energy spectra in Fig.\ \ref{Espectrum}; the presence of the
impurity potential enlarges the energy gap around the chemical potential
$\mu$. Instead of a spatially fairly extended state just below $\mu$
there is now a localized state at lowered energy.
%
%
%
%
Again, with increasing excitation strength $V_t$ we see a red shift of the
resonance in the parallel oscillations. At strong excitation we can not
refer back to the Kohn-Sham spectra of the ground state, but we can 
expect that the energy pumped into the system makes the states just 
below the gap more extended, thus reducing the effective energy gap.
The perpendicular oscillation in Fig.\ \ref{Piy}a is more complex
than before. Again, there is a resonance at the same energy as 
in the parallel mode that is slightly red shifted with increased
excitation, but there now appear resonances at higher energies indicating
the presence of more localized modes in the system, an extra excitation
mode located at the impurity site. This is again supported by a look
at the Kohn-Sham energy spectra in Fig.\ \ref{Espectrum}a, at the
changes just above the gap caused by the impurity.   

\begin{figure}[htbp!] 
\begin{center}
\includegraphics[bb=152 215 430 615,clip,width=8.0cm]{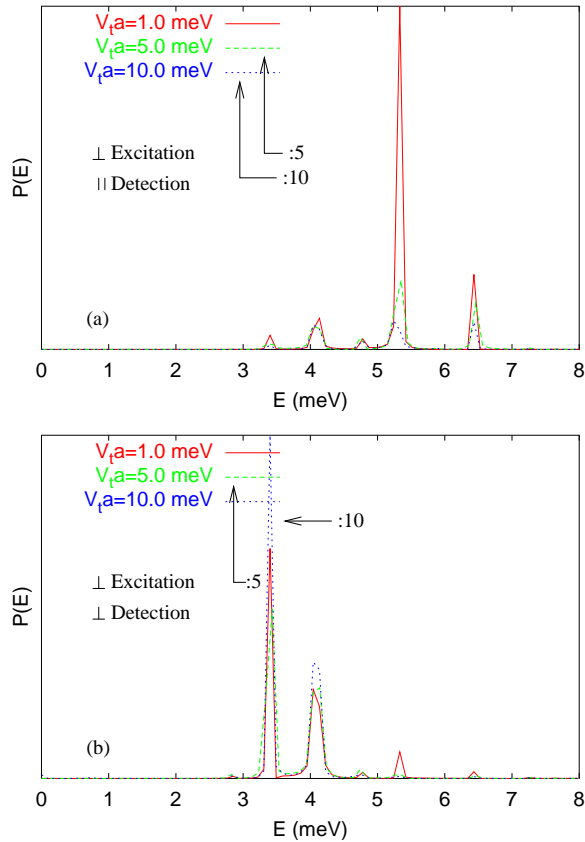}
\end{center}
\caption{Color online.
         The Fourier power spectrum of the time series for the
         expectation values of the coordinates 
         perpendicular ($\perp$) (a),
         and parallel ($\parallel$) (b) to the long axis 
         of the system. The excitation is {\bf\em perpendicular} to the same
         axis. Division factors for the stronger excitations
         are indicated in the figure. {\bf\em Impurity is present}
         in the system. $B=0$ T, $T=1$ K, and $N=10$. }
\label{Pi_an}
\end{figure}

\begin{figure}[htbp!] 
\begin{center}
\includegraphics[bb=152 215 430 615,clip,width=8.0cm]{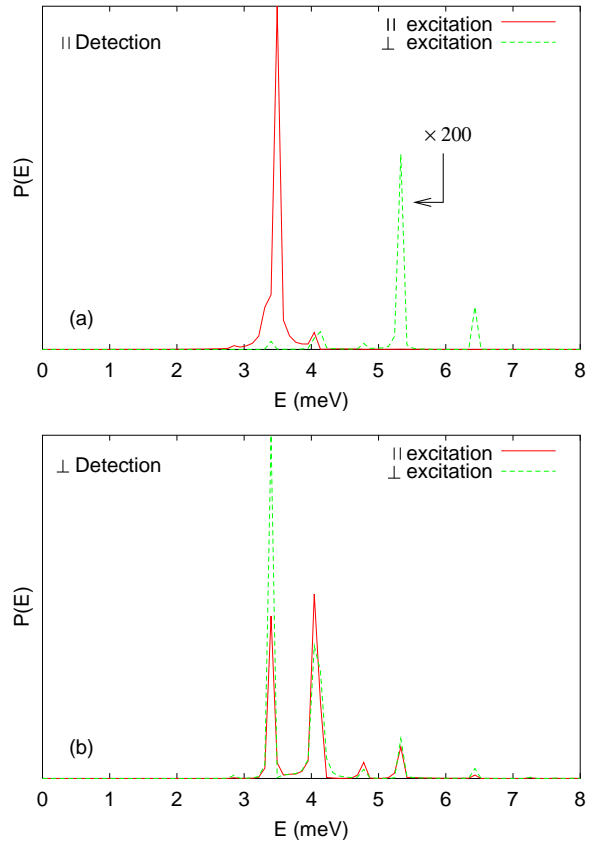}
\end{center}
\caption{Color online.
         Comparison of the Fourier power spectrum of the time series 
         for the expectation values of the coordinates 
         parallel ($\parallel$) (a),
         and perpendicular ($\perp$) (b) to the long axis 
         of the system in the case of an excitation 
         {\bf\em perpendicular} or {\bf\em parallel} to the same
         axis. A multiplication factor is indicated in the figure. 
         {\bf\em Impurity is present} in the system. $V_ta=1.0$ meV, 
         $B=0$ T, $T=1$ K, and $N=10$.}
\label{Fig_Py_og_an}
\end{figure}

The case of the initial excitation pulse aligned perpendicular with the
long axis is review in Fig.\ \ref{Pi_an}. We see that again in the
$\perp$ detection we have the peak structure encountered in Fig.\ \ref{Piy}a
as could be expected, but the slight presence of a $\parallel$ excitation
in the initial pulse is not of the dipole kind so mostly higher order
modes are excited in the $\parallel$ direction. 
%
%
%
%
Exactly this, is further
displayed in Fig.\ \ref{Fig_Py_og_an}a, where, in addition, the relative 
strength of the response is also visible. In contrast, the strong confinement
in the perpendicular direction of the system or the wire ensures the 
$\perp$ response to be not strongly dependent on the main 
excitation polarization, see Fig.\ \ref{Fig_Py_og_an}b. 
%
%
%
%

\begin{figure}[htbp!] 
\begin{center}
\includegraphics[width=8.0cm]{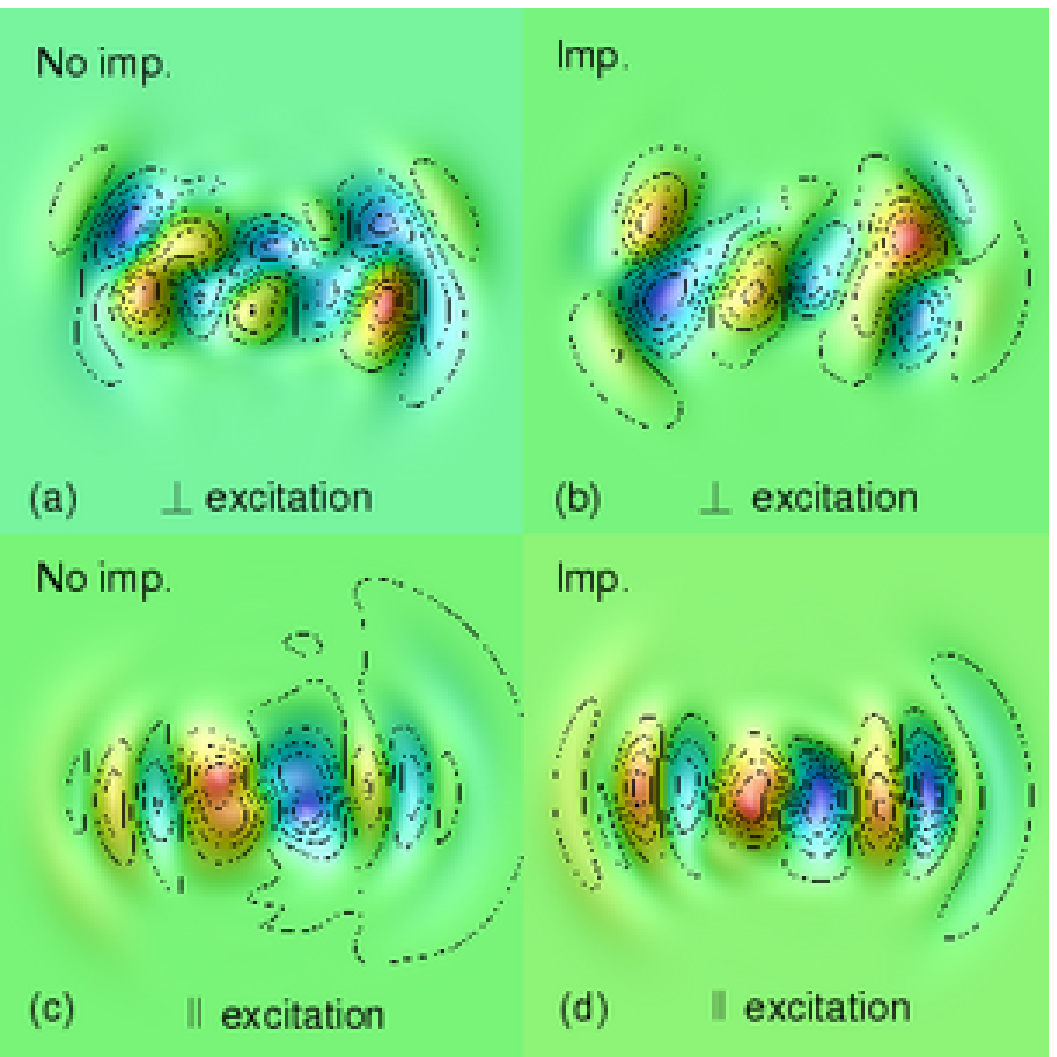}
\end{center}
\caption{Color online.
         The induced density at time $t=50$ ps for $\perp$ excitation
         for a system without (a), and with an impurity (b),
         and for $\parallel$ excitation without (c), and with an 
         impurity (d). The excitation strength 
         $V_ta=1.0$ meV (a), $5.0$ meV (b)
         $10^{-2}$ meV (c), $1.0$ meV (d). $B=0$ T, $T=1$ K, and $N=10$.}
\label{I_zz}
\end{figure}

To clearly demonstrate the collective oscillations in the system we show in
Fig.\ \ref{I_zz} the induced electron density for the system with and without
an embedded impurity for the two polarizations of the excitation pulse.
The two figures Fig.\ \ref{I_zz}a and b for the 
$\perp$ excitation show that the ends of the
system oscillate in the $\perp$ direction while the oscillation around the
impurity is at least at this moment parallel to the long axis. In case
of no impurity in the system the oscillation of the center part is more
perpendicular. When the excitation is applied $\parallel$ to the system
(see Figs.\ \ref{I_zz}c and d) the main oscillation mode is also along the
system of the fundamental dipole type as can be seen by comparing the induced
density to the total density in Fig.\ \ref{Density}. In this case the
oscillation seems to be more extended to the whole system when there is an
impurity present in it, but when there is no impurity present the center of the
system displays a tiny bit of a $\perp$ mode superimposed on its motion.
In fact, the induced density can show higher order oscillation modes that
we can not discern with our time series of the expectations values for the
the coordinates $x$ and $y$, which are most sensitive to dipole oscillations
occuring in the system.\cite{puente99:3266,Puente01:235324}    
%
%
%
%

\begin{figure}[htbp!] 
\begin{center}
\includegraphics[bb=152 115 430 715,clip,width=8.0cm]{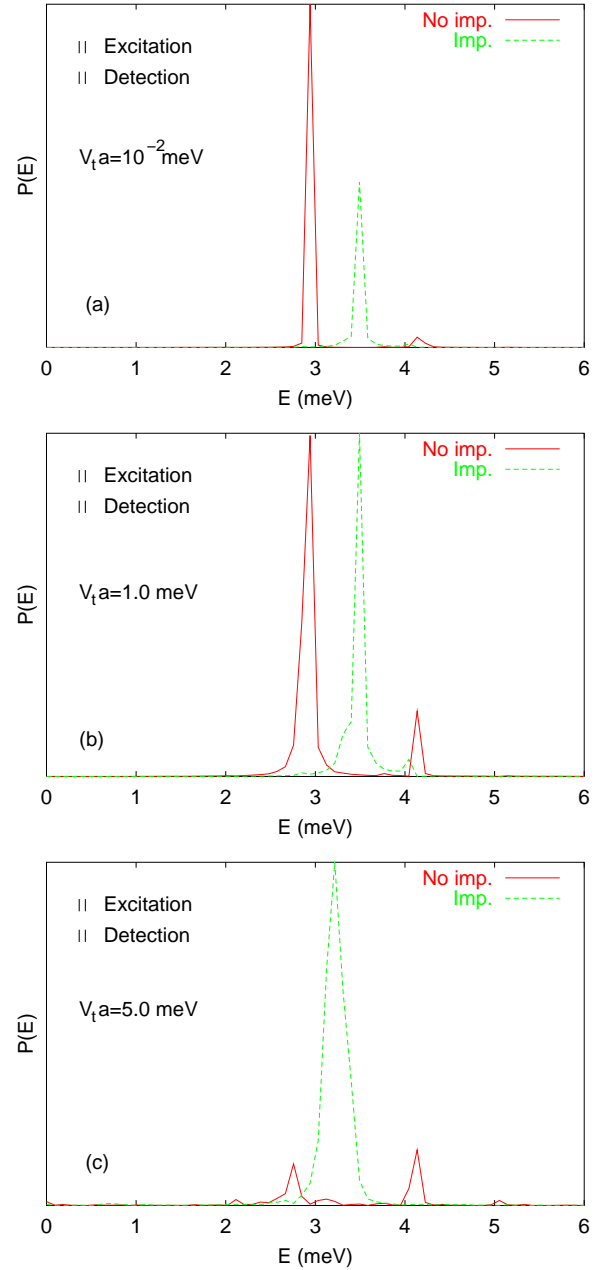}
\end{center}
\caption{Color online.
         The Fourier power spectrum of the time series for the
         expectation values of the coordinate 
         parallel ($\parallel$) to the long axis 
         of the system with or without an impurity 
         in the case of an excitation along the same axis.
         $B=0$ T, $T=1$ K, and $N=10$.}
\label{Pi0_ycomp_x}
\end{figure}

The red shift of the collective resonance for the system with an embedded
impurity is further explored in Fig.\ \ref{Pi0_ycomp_x} in comparison with
the response of the system without an impurity. Clearly, the resonance is
red shifted and broadened with increasing excitation strength $V_t$ pointing
towards a reduction in the binding energy of the impurity. 

Comparison between the sub figures of Fig.\ \ref{Pi0_ycomp_x} reveals
that with increasing excitation strength more energy flows into a 
simple dipole oscillation of the system when an impurity is present in it.  
%
%
%
%
This is confirmed by the
induced electron density in this case showing large oscillations
of the electron density around the impurity site parallel to the system, 
and mostly in harmony with the oscillations in the rest of the system, reflected
in a single resonance. The broadening of the resonance
is indicative of a finer structure that we are unable to resolve in our 
calculations.

\begin{figure}[htbp!] 
\begin{center}
\includegraphics[bb=152 115 430 715,clip,width=8.0cm]{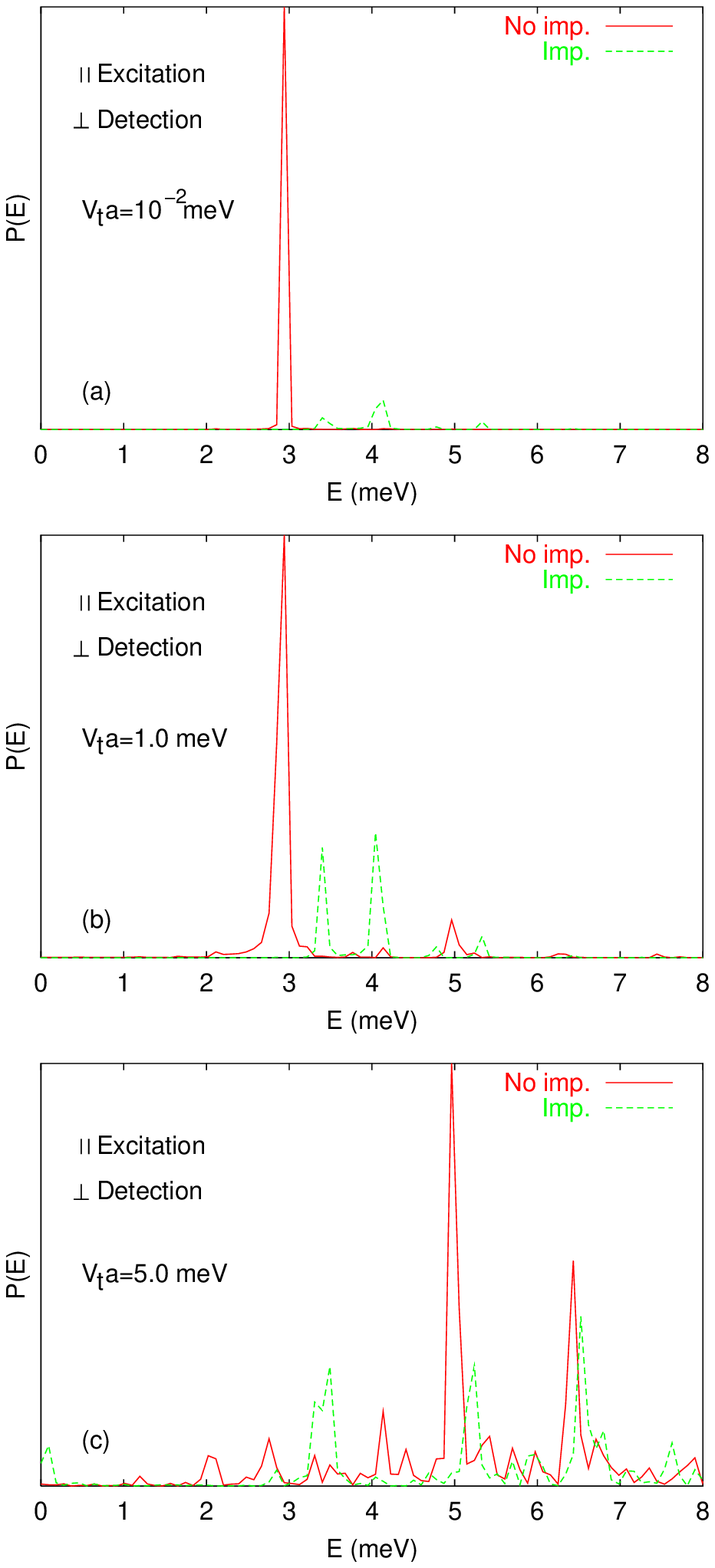}
\end{center}
\caption{Color online.
         The Fourier power spectrum of the time series for the
         expectation values of the coordinate 
         perpendicular ($\perp$) to the long axis 
         of the system with or without an impurity 
         in the case of an excitation along the same axis.
         $B=0$ T, $T=1$ K, and $N=10$.}
\label{Pi0_ycomp_y}
\end{figure}

The case is different with the perpendicular oscillation
shown in Fig.\ \ref{Pi0_ycomp_y}, where in the end a strong excitation will
lead to many modes becoming active. So, even though the parallel 
collective oscillation remains simple the perpendicular one is generally
more complex.
%
%
%
%

\subsection{The $B=1$ T case}
\begin{figure}[htbp!] 
\begin{center}
\includegraphics[bb=152 215 430 615,clip,width=8.0cm]{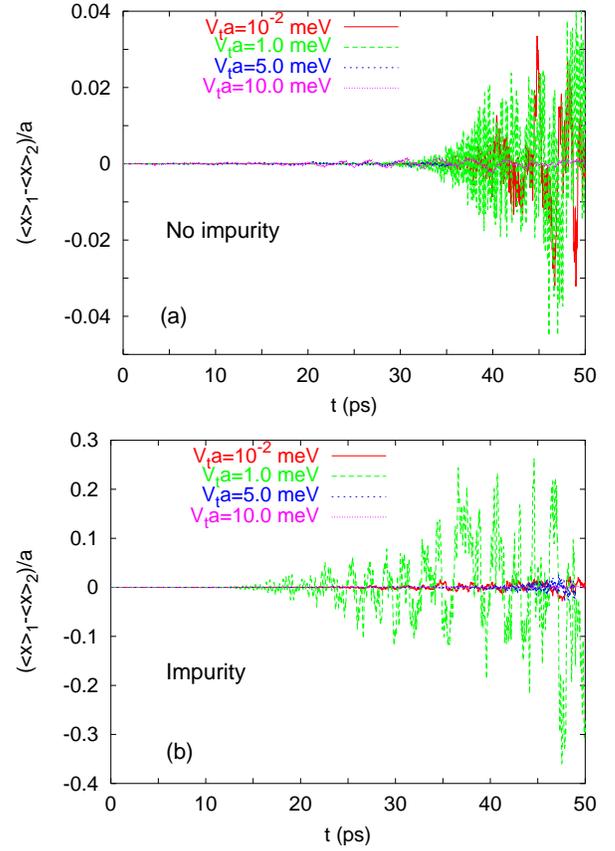}
\end{center}
\caption{Color online.
         The time evolution of spin oscillations, characterized by
         $(\langle x\rangle_\downarrow - \langle x\rangle_\uparrow)/a$,
         for a system without (a) and with (b) an impurity.
         $\Delta t=0.0025$ ps, $B=1$ T, $T=1$ K, and $N=10$.}
\label{Si0_x}
\end{figure}

The collective oscillations in the finite magnetic field $B=1$ T
are for short times after the initial excitation pulse very similar
to the oscillations encountered in the system in no magnetic field
as could be guessed by comparing the Kohn-Sham energy spectra for
the two cases, see Fig.\ \ref{Espectrum}. It has though to be kept in
mind that the Lorentz force now ensures a quick communication between
the parallel and the perpendicular modes, they are not independent.
But more happens: The finite size of the system together with the
small Zeeman energy, and the finite temperature $T=1$ K establish a
tiny spin polarization in the ground state, i.e.\ the two spin densities
do not have exactly the same spatial dependence. The spatially varying
exchange force in the LSDA gives a different restoring force to the
spin densities. All these factors lead to a slow flow of energy into
the individual oscillations of the spin densities as can be seen in 
Fig.\ \ref{Si0_x}. This flow is so slow that after 50 ps the system
has still not entered a ``steady state'' as in the case of $B=0$ T.
%
%
%
%
We are still observing switch-on effects for the collective
oscillations of the spin densities after 50 ps. 
In the spin oscillation we see present both high and low frequency
components but the lack of a steady state prevents us from making
a Fourier analysis. Just to avoid any misconception we should mention
that the time increment of $\Delta t=0.0025$ ps gives us 20,000 points
in the range of 50 ps and at a finer level of resolution the spin 
oscillation looks very smooth with no apparent noise like might be
hastily inferred by looking at Fig.\ \ref{Si0_x}.    

Even more amazing is the fact that to begin with this spin oscillation
increases with the strength of the excitation $V_t$ but, as is seen
in Fig.\ \ref{Si0_x}, for a still stronger excitation it again
{\em decreases}. Stronger excitation of the system prevents the energy
in flowing into the collective spin oscillations, but directs it to
the center-of-mass motion and the other collective charge oscillations.
This behavior is definitely a nonlinear phenomena that can not
be observed in a model of linear response.
%
%
\begin{figure}[htbp!] 
\begin{center}
\includegraphics[bb=60 68 380 250,clip,width=8.0cm]{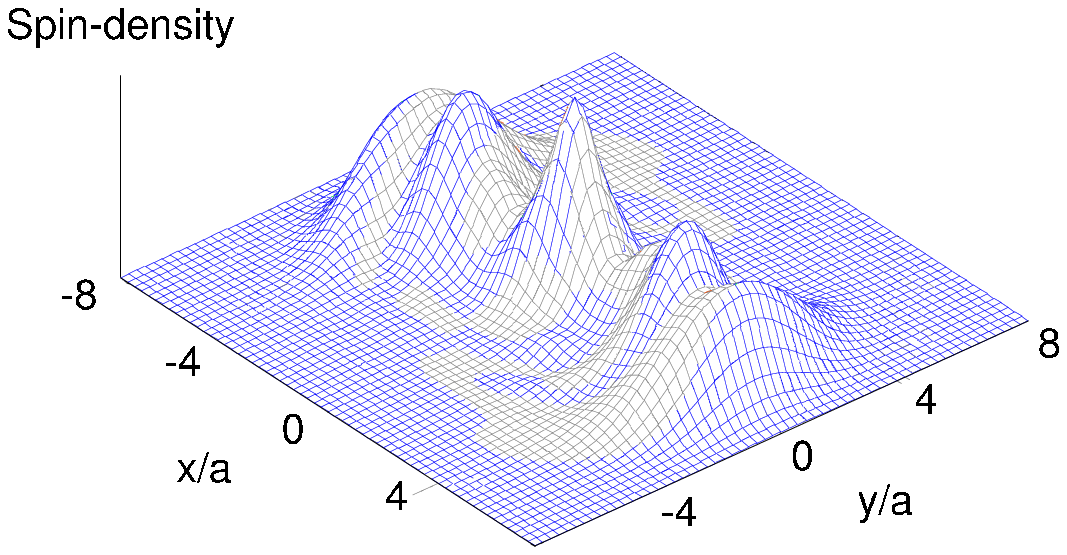}
\end{center}
\caption{Color online.
         The two components of the spin density at the
         time point $t=50$ ps in a system with an impurity.
         The spin densities are very similar in shape and are
         superimposed here on one another with different shades.  
         $V_ta=10^{-2}$ meV, $B=1$ T, $T=1$ K, and $N=10$.}
\label{Spin_density}
\end{figure}
%
%

The presence of the impurity also quite effectively directs more
energy into the spin oscillation as is seen in Fig.\ \ref{Si0_x}.
This can be explained by a stronger spatial variation of the 
exchange force around the impurity.        

The regularity of the collective spin oscillations can be seen in
Fig.\ \ref{Spin_density} where simply both spin densities are
seen almost overlapping at $t=50$ ps. The peak structure of the
spin densities closely follows the structure of the 
ground state charge density displayed in Fig.\ \ref{Density}.  
\section{Discussion and summary}     

Without the external magnetic field present the collective charge
oscillations excited in the system by the short THz pulse do reach
a ``steady state'' condition within 1 or 2 picoseconds 
after the excitation is switched off allowing a Fourier
analyzes to discern the modes present. At low excitation level we
observe a linear regime, but at higher excitation nonlinear effects
show up. The excitation gap in the system is reduced indicating 
a less localized single-electron states and in particular the 
effective binding energy of the impurity is reduced. 

At finite magnetic field ($B=1$ T) a slowly increasing collective oscillation
of the spin densities has not reached a ``steady state'' condition after
50 ps. As the spin oscillation is directly coupled with the 
collective charge oscillations a Fourier analyses is not possible in this
case for either component. 
The excitation of the collective spin oscillations
is much more effective when an impurity is present in the system, since then
the initial spin densities at $t=0$ have a larger spatial variation. 
The more amazing fact is that with increased intensity of the initial
excitation the onset of the spin oscillation can be effectively shut
down. We contribute this to the flow of the energy nonadiabatially pumped 
into the system being more easily directed to the oscillations of the
charge density in this case.   

For the results shown in this paper we have always used the same
type of excitation pulse. What would happen if we use a longer
pulse with several oscillations or of a different form. We can not
claim that we have thoroughly investigated this question, but our limited
experimenting with this seems to indicate that the pulse form is indeed
important for the heights of the resonance peaks, but does not affect
their location strongly. We are pumping energy into the system by the excitation
in varying amounts depending on the peak shape and duration and it 
can be expected that this also affects the initial distribution of the
energy between the modes. By removing the system further from equilibrium
than we do we are sure that stronger nonlinear effects can occur 
with considerable change of peak locations also. We already see the
precursors of this in the redshift of the peaks connected to the
impurity embedded in the system.     

We are aware that our choice of an impurity potential represented by a
Gaussian well does not correspond to unscreened Coulomb impurities 
that might be situated in the plane of the 2DEG in a nanosystems. 
This form has only been selected in order to keep the 
truncation of the Fock-Darwin basis within reasonable bounds in order
to enable the ``long-time'' integration of the time evolution operator
to extend to 50 ps with 20,000 steps, within each one the self-consistency
was attained with further 4 loops to properly update the density operator.
With these requirements in mind it should be reiterated that the present
model was not designed to describe any known experimental device, but
rather it was constructed to give qualitative insight into time-dependent
phenomena that can take place in a nanosystem that is irradiated by
short and intensive THz pulse, in response to the preparations underway
in several experimental groups to study these phenomena. 
We only study the behavior of the system for a short time after the
application of the excitation and neglect all possible damping and
dissipation effects that would be caused by the coupling of the system
to its immediate environment. These are certainly very important 
effects that await further investigation.      

%
%
\begin{acknowledgments}
      The research was partly funded by the Icelandic Natural Science Foundation,
      the University of Iceland Research Fund, the National Science 
      Council of Taiwan under Grant No.\ 91-2119-M-007-004, and the 
      National Center for Theoretical Sciences, Tsing Hua University, Hsinchu 
      Taiwan. VG acknowledges instructive discussions with Llorens Serra and 
      a beneficial exchange of programming ideas with Ingibj{\"o}rg 
      Magn{\'u}sd{\'o}ttir and Gabriel Vasile. 
\end{acknowledgments}
%
%

\appendix
\section*{The generalized one-electron overlap integral}
In order to reduce the memory size needed to store the generalized
overlap integral it is convenient to split it into two parts by
applying the completeness of the basis once
\begin{equation}
      \Phi_{\alpha pq\beta} = \int d\mathbf{r}\;\phi_\alpha^*(\mathbf{r})
      \phi_p^*(\mathbf{r})\phi_q(\mathbf{r})\phi_\beta (\mathbf{r})
      = \sum_\delta \Phi_{\alpha p\delta}\Phi_{q\beta\delta}, 
\end{equation}
where an analytic expression for the wave functions corresponding to the
Fock-Darwin basis 
states $|\alpha\rangle =|M_\alpha ,n_\alpha\rangle$ gives
\begin{eqnarray}
      \Phi_{\alpha p\delta} = 2\pi a^2\; \delta_{M_\alpha +M_p,M_\delta}
      \; \beta_\alpha\beta_p\beta_\delta   \;                \nonumber
      2^{(|M_\alpha|+|M_\beta|+|M_\delta|)/2}  \\ \nonumber
      \sum_{i_\alpha =0}^{n_\alpha} \sum_{i_p=0}^{n_p} 
      \sum_{i_\delta =0}^{n_\delta} \frac{(-1)^{i_\alpha +i_p+i_\delta}}
      {i_\alpha !i_p !i_\delta !}
      \left({n_\alpha +|M_\alpha |\atop n_\alpha -i_\alpha}\right)
      \left({n_p      +|M_p      |\atop n_p      -i_p     }\right) \\ \nonumber
      \left({n_\delta +|M_\delta |\atop n_\delta -i_\delta}\right)
      \left( \frac{|M_\alpha|+|M_\beta|+|M_\delta|}{2} 
      +i_\alpha +i_p+i_\delta \right)!       \\ \nonumber        
      \times {\left(\frac {2}{3}\right)}^{[(|M_\alpha|+|M_\beta|+|M_\delta|)/2] 
      +i_\alpha +i_p+i_\delta +1},
\end{eqnarray}
with 
\begin{equation}
      \beta_\alpha = \frac{1}{2^{(|M_\alpha|+1)/2}} 
      \sqrt{ \frac{n_\alpha !}{\pi (|M_\alpha |+n_\alpha )}} . 
\end{equation}
\bibliographystyle{apsrev}
\bibliography{/home/vidar/Textar/Greinar/Nano_QD/mod_qd}

%
\end{document}